# Charged impurity scattering in bilayer graphene

Shudong Xiao[1], Jian-Hao Chen[1,2], Shaffique Adam[1,3], Ellen D. Williams[1,2] and Michael S. Fuhrer[1,2*]

*[1]Center for Nanophysics and Advanced Materials, [2]Materials Research Science and Engineering Center, and [3]Condensed Matter Theory Center, Department of Physics, University of Maryland, College Park, MD 20742 USA*

We have examined the impact of charged impurity scattering on charge carrier transport in bilayer graphene (BLG) by deposition of potassium in ultra-high vacuum at low temperature. Charged impurity scattering gives a conductivity which is supra-linear in carrier density, with a magnitude similar to single-layer graphene for the measured range of carrier densities of 2-4 x $10^{12}$ $cm^{-2}$. Upon addition of charged impurities of concentration $n_{imp}$, the minimum conductivity $\sigma_{min}$ decreases proportional to $n_{imp}^{-1/2}$, while the electron and hole puddle carrier density increases proportional to $n_{imp}^{1/2}$. These results for the intentional deposition of potassium on BLG are in good agreement with theoretical predictions for charged impurity scattering. However, our results also suggest that charged impurity scattering alone cannot explain the observed transport properties of pristine BLG on $SiO_2$ before potassium doping.

*email: mfuhrer@umd.edu*

Bilayer graphene (BLG)[1,2] is a unique electronic material distinct from single-layer graphene (SLG)[3]: while SLG has a massless, gapless electronic dispersion $E(k) = \pm\hbar v_F|k|$, BLG has a low-energy dispersion which is approximated[2,4] by massive valence and conduction bands with zero gap: $E(k) = \pm\hbar^2k^2/2m^*$, where the effective mass is $m^* = \gamma_1/2v_F^2$, with $\gamma_1 \approx 0.39$ eV the interlayer hopping matrix element, $v_F \approx 1.1 \times 10^6$ m/s the Fermi velocity in single layer graphene, and $\hbar$ Planck's constant. BLG has attracted interest because of a tunable bandgap[5,6,7,8], and unusual quantum Hall physics with an eight-fold degenerate zero energy Landau level[1]. However, little is known about disorder and charge-carrier scattering in BLG. Similar to SLG, linear $\sigma(n)$ is observed experimentally[9], with mobilities limited to $<10^4$ cm$^2$/Vs. However, unlike SLG, linear $\sigma(n)$ is expected for both charged impurities and short-range scatterers within the complete screening approximation[10,11,12], hence the dominant disorder scattering mechanism in BLG remains an open question.

SLG provides a starting point for understanding the effects of disorder in BLG. In SLG on $SiO_2$ substrates[13] impurity scattering is dominated by charged impurities with a typical density $n_{imp}$ of a few $10^{11}$ cm$^{-2}$, which give rise to a linear conductivity as a function of charge carrier density, i.e. $\sigma(n) = ne\mu$ [14,15] with constant mobility $\mu$, with additional contributions from weak short-range scatterers with $\sigma(n) \sim$ constant[16]. At low $n$, the random potential from charged impurities produces electron and hole puddles with a characteristic carrier density $n^*$, giving rise to a minimum conductivity $\sigma_{min} = n^*e\mu \approx (4\text{-}10)e^2/h$. To leading order, $n^* \propto n_{imp}$ and $\mu \propto n_{imp}^{-1}$, so $\sigma_{min}$ varies only weakly with $n_{imp}$[14,17]. Charged impurities have been predicted to lead to stronger scattering in BLG compared to SLG[10], consistent with the generally lower mobilities observed for BLG compared to SLG. However, as we discuss below, this prediction was based on two severe approximations for the bilayer case (complete screening, and zero impurity-graphene distance) and a more complete treatment indicates that BLG and SLG should have similar mobility for a similar density of charged impurity scatterers. In contrast to SLG, the random charged impurity potential in BLG is well-screened, and $n^* = (n_{imp}/\xi^2)^{1/2}$, i.e. $n^*$ is simply the fluctuation in the impurity number within an area given by the square of the puddle correlation length $\xi$. This leads to a strong prediction for the variation of the minimum conductivity on the density of trapped charges $\sigma_{min} \propto n_{imp}^{-1/2}$ which can be tested experimentally.

Here we experimentally measure the scattering rate for charged impurities on BLG by depositing potassium on BLG in ultra-high vacuum (UHV) at low temperature. Charged impurity scattering gives a carrier-density-dependent conductivity $\sigma(n)$ which is supralinear in $n$, with similar magnitude to single layer graphene for the measured range of carrier densities of 2-4 x $10^{12}$ $cm^{-2}$. The conductivity is in good agreement with that calculated within the Thomas-Fermi (TF) screening approximation[10] once the finite screening length and impurity-graphene distance are taken into account. The dependence of the minimum conductivity and the residual carrier density on charged impurity density are well-described by $\sigma_{min} \propto n_{imp}^{-1/2}$ and $n^* = (n_{imp}/\xi^2)^{1/2}$ in agreement with theoretical expectations, though the puddle correlation length $\xi$ is significantly larger than predicted theoretically. However, the experimentally measured magnitude and carrier-density dependence for charged impurity scattering on BLG indicates that unlike SLG, charged impurities alone cannot explain the observed transport behavior of pristine BLG samples on $SiO_2$, i.e. before the intentional addition of charged impurities. We infer the presence of an additional source of disorder in the undoped BLG that gives rise to $\sigma(n) \sim n$.

BLG is mechanically exfoliated from Kish graphite onto 300 nm $SiO_2$ on Si substrates. Figure 1a shows the BLG device used in this work. The Au/Cr electrodes are defined by electron-beam lithography and the doped Si acts as a back gate. Figure 1b shows the micro-Raman spectrum measured for this device. The Raman G' band can be fit with four Lorentzian components (Fig. 1b); their relative peak positions and magnitudes are similar to those in Ref [18], and are indicative of Bernal stacking. After annealing in $H_2$/Ar at 400 ºC[19], the device was mounted on a cold finger in UHV chamber and an overnight bakeout was performed in vacuum. In UHV, the charged-impurity density $n_{imp}$ was varied systematically by deposition of potassium atoms from a controlled source at a sample temperature $T$ = 10K. Conductivity as a function of gate voltage $\sigma(V_g)$ was measured *in situ* at different K concentrations; the carrier concentration is given by $n = (c_g/e)(V_g - V_{g,min}) = [7.2 \times 10^{10}\ cm^{-2}V^{-1}](V_g - V_{g,min})$ with $c_g = 1.15 \times 10^{-8}$ F/cm$^2$ the gate capacitance per unit area and $V_{g,min}$ the gate voltage of minimum conductivity.

Figure 2a shows $\sigma(V_g)$ measured at different K doses for BLG and, for comparison, Figure 2b shows similar data for SLG taken from Ref. [14]. Before K doping, the annealed BLG sample has a lower mobility (1,200 cm$^2$/Vs) than pristine SLG prepared similarly (13,000 cm$^2$/Vs). This is typical for our $H_2$/Ar annealed BLG samples, which show mobility 2-5 times lower than un-annealed BLG, and ~10 times lower than SLG devices on the same $SiO_2$ substrates (annealing SLG does not appreciably change the mobility). K doping shifts the transport curve to the negative gate voltage side, lowers the mobility, decreases $\sigma_{min}$, broadens the minimum conductivity plateau and makes the $\sigma(V_g)$ curve nonlinear.

For uncorrelated impurities, the mobility is inversely proportional to the impurity density $\mu = \frac{C}{n_{imp}}$. The non-linearity of $\sigma(V_g)$ indicates that mobility, and thus $C$, is a function of carrier density, unlike SLG where $C$ is a constant. To quantify our results we introduce an initial impurity density $n_{imp,0}$, so that the total impurity density is $n_{imp} = n_{imp,0} + n_K$, where $n_K$ is the potassium concentration. While the charged impurities corresponding to $n_{imp,0}$ could in principle have opposite charge or be at a different distance from the bilayer graphene sheet than $n_K$, to avoid introducing too many parameters, and consistent with results from residual impurities on single-layer graphene [16], we assume $\frac{1}{\mu(n)} = \frac{n_{imp,0}}{C(n)} + \frac{n_K}{C(n)}$. We assume that $n_K$ is given by the shift of $V_{g,min}$, i.e. $n_K = (c_g/e)\Delta V_{g,min}$ which is exact within the parabolic approximation for the BLG Hamiltonian[10]; below we show that for the range of densities we consider, this approximation remains very good for the hyperbolic Hamiltonian.

Figure 3a shows the inverse electron mobility $1/\mu$ as a function of $n_K$ at $V_g = 30$ V and 60 V for BLG. $1/\mu$ vs. $n_K$ is linear as expected, and we determine $C(n)$ as the inverse of the slope of $1/\mu$ vs. $n_K$, yielding $C(60\ \text{V}) = 5.1 \times 10^{15}\ \text{V}^{-1}\text{s}^{-1}$ and $C(30\text{V}) = 4.2 \times 10^{15}\ \text{V}^{-1}\text{s}^{-1}$. For $V_g < 30$ V the measurement is influenced by the minimum conductivity region, and $1/\mu$ vs. $n_K$ is not linear, so $C$ could not be extracted. For BLG, $n_{imp,0}$ varies systematically from $3.4 \times 10^{16}\text{m}^{-2}$ at $V_g = 30$ V to $4.3 \times 10^{16}\text{m}^{-2}$ at $V_g = 60$ V. We find that the initial impurity density $n_{imp,0}$ for BLG is one order of magnitude higher than for SLG (see discussion below), the data for which are shown for comparison

Figure 3b shows the complete measured dependence of $C(V_g)$ for BLG (solid squares). Data from a second sample is also shown (solid circles), with similar results. For comparison, the SLG value, $C = 5 \times 10^{15}$ V$^{-1}$s$^{-1}$ [14] is shown in blue. The similarity to the values for BLG indicates that the scattering cross section for charged impurities in BLG is very similar to SLG. The black line shows the previously calculated result[10,20] for $C(V_g)$ within the complete screening approximation with $d = 0$. The red and purple lines show $C(V_g)$ calculated within the Thomas-Fermi (TF) approximation without making the complete-screening approximation[10] for impurity-graphene distances $d = 0$ (red) and $d = 0.43$ nm (purple; the expected potassium-graphene distance of 0.26 nm[21] plus one-half the interlayer separation of 0.34 nm). The experimental data are close to the TF calculation with somewhat smaller magnitude and less carrier density dependence[22,23,24].

Figure 4 shows $\sigma_{min}$ as a function of $n_K$. The minimum conductivity decreases with increasing charged impurity concentration. The residual carrier density $n^* = \sigma_{min}/e\mu = \sigma_{min}(n_{imp,0} + n_K)/eC$. Since we do not know the mobility at $V_g = 0$, we use $C(30\text{ V}) = 4.2 \times 10^{15}$ V$^{-1}$s$^{-1}$ and $n_{imp,0}(30\text{ V}) = 3.4 \times 10^{16}$ m$^{-2}$ to estimate $\mu = C/(n_{imp,0} + n_K)$. Figure 4 shows $n^*$ as a function of $n_K$. $n^*$ increases with charged impurity doping, as expected. The solid lines in Figure 4 show fits to the theoretically predicted behavior $n^* = [(n_{imp,0} + n_K)/\xi^2]^{1/2}$ and $\sigma_{min} = Ce[(n_{imp,0} + n_K)\xi^2]^{-1/2}$. The only free parameter $\xi$ is found to be 32 nm. This is significantly larger than the correlation length $\xi = 9$ nm calculated within the self-consistent model using TF screening. We likely overestimated $C$ by as much as a factor of 3 in using $C(30\text{ V})$ (see Figure 3b), and therefore $\xi$ may be as much as 40% smaller (~18 nm), but still twice the calculated value. A similar discrepancy (self-consistent theory overestimating $n^*$) is found in SLG[14].

The theoretical results discussed above rely on the parabolic approximation for the dispersion relation for BLG [2], only valid for carrier densities much lower than $n_0 = (v_F m^*/\hbar)^2/\pi$ ~ 2 × $10^{12}$ $cm^{-2}$. The experimental results presented here cross over from this low density limit to much higher densities where the parabolic approximation for the Hamiltonian breaks down. We briefly examine the robustness of the theoretical results for BLG transport at low density[10] to the situation when the carrier density (or equivalently, the impurity density) is much larger than $n_0$. Our main finding is that the results for higher density are qualitatively very similar to those found using the parabolic approximation. The crossover Hamiltonian reads [2]

$H = \sigma_x \otimes [v_F(\sigma_x, \sigma_y)\cdot \hbar\vec{k}] + [(I_2 - \sigma_z)/2] \otimes \gamma_1 \sigma_x$, where $I_2$ is the identity matrix and $\sigma_{x,y,z}$ are the Pauli matrices. The dispersion relation is hyperbolic, with $E_b = \hbar^2 k^2/(2m^*)$ and $E_s = \hbar v_F |k|$ as the low density and high density asymptotes, where $v_F = 1.1 \times 10^6 m/s$ is the SLG Fermi velocity and $m^* = \gamma_1/(2v_F^2) \approx 0.033 m_e$ is the low density effective mass for BLG. Analogous to the treatment in Ref[25] for SLG, for the crossover Hamiltonian the scattering time reads

$$\frac{\hbar}{\tau(\varepsilon(k))} = 2\pi \sum_{k'} n_{imp} \left| \frac{v(q,d)}{\varepsilon(q)} \right|^2 F(\theta)(1-\cos\theta)\delta(\varepsilon(k)-\varepsilon(k')), \qquad (1)$$

where the wavefunction overlap $F(\theta) = (1/4)[1-\eta+(1+\eta)\cos\theta]^2$, and $\eta = (1+n/n_0)^{-1/2}$ parameterizes the crossover. Within TF, the dielectric function $\varepsilon(q,n) = 1 + v(q)\nu(n)$, and density of states $\nu(n) = (2m^*/\pi\hbar)\sqrt{1+n/n_0}$ . The mobility calculated using Eq. 1 is shown in Fig. 3b (green line). As seen in the figure, while the modified Hamiltonian gives a slightly larger mobility, it is not significantly different from the low density parabolic dispersion approximation.

We can also examine the transport properties at low density, close to the Dirac point. Applying the self-consistent transport theory [17] to the parabolic approximation for bilayer graphene [10] gives $\bar{n} \equiv \left(\frac{c_g}{e}\right) V_{g,\min} = n_{imp}$ and $n^* = [n_{imp}/\xi^2]^{1/2}$. Using the crossover Hamitonian we find $n^* \approx n_{imp} C_0\left[\alpha\sqrt{n^*/n_0}\right] + \sqrt{4 n_{imp} n_0 C_0[\alpha\sqrt{n^*/n_0}]}$, where $\alpha = 4d\sqrt{2\pi n_0}$ and $C_0[x] = \partial_x\left[x e^x \int_x^\infty t^{-1} e^{-t} dt\right]$. The numerical solution for the electron and hole puddle density using the crossover Hamiltonian is remarkably close to the parabolic result $n^* = [n_{imp}/\xi^2]^{1/2}$ with only about a five percent decrease in the value of $\xi$. The correction to $\bar{n}$ is more significant

$$\frac{\bar{n}}{n_{imp}} \approx \left[\frac{n_{imp}/n_0 + 4\sqrt{1+\sqrt{\beta n_{imp}}}}{4(1+\sqrt{\beta n_{imp}})}\right], \tag{2}$$

where $\beta = 1/\xi^2 n_0^2 \approx 6.4\times10^{-13}\,cm^2$, and the right hand side of Eq. 2 changes from unity at low impurity density to about 0.8 for the highest impurity densities we consider. This indicates that we may have underestimated the impurity concentration from $n_{imp} = \bar{n}$ by up to ~20%, which would indicate that $C(n)$ may be higher than shown in Fig. 3b by up to ~20%.

Overall, the magnitude and carrier-density dependence of $C$ and the impurity density dependence of $n^*$ and $\sigma_{min}$ are in good qualitative agreement with the theory of charged impurity scattering in BLG. However, $C$ is somewhat smaller, and $\xi$ somewhat larger, than expected theoretically, which both indicate that screening is not as effective as predicted. A possible explanation is the opening of a gap at the Dirac point in biased bilayer graphene[5,6,7,8], which we have not treated theoretically. The reduced screening in gapped BLG has also been put forth to explain the dependence of flicker noise on gate voltage in BLG[26]. From the optical measurements of Ref. [7] we can estimate that the maximum band-gap we would induce is about 100 meV, while for $n_{\mathrm{imp}}$ = $5.3 \times 10^{12}\mathrm{cm}^{-2}$ , we can estimate the unscreened disorder potential to be about 200 meV. One can expect that the signatures in transport experiments of the electric field induced band-gap to be negligible when the disorder potential fluctuation is much larger than the band-gap [27]. We expect the opening of a bandgap in BLG to have an even smaller effect on transport in the high-density regime (data in Figure 3b), where the change in density of states at the Fermi energy in BLG induced by gap opening is estimated to be a few percent.

Lastly, we discuss the nature of scattering in BLG on $SiO_2$. Our experimental finding that the magnitude of charged-impurity scattering in BLG is similar to SLG is surprising given that pristine BLG typically shows lower mobility (~10 times for our $H_2$/Ar annealed samples) than SLG on nominally identical $SiO_2$ substrates. We note that the $H_2$/Ar annealing process itself significantly lowers the mobility of BLG without affecting SLG, which is not understood. The variation of $C$ with $V_g$ is also inconsistent with the linear $\sigma(V_g)$ observed in BLG[9]. Together, these observations indicate that another source of disorder may dominate BLG on $SiO_2$. This may be consistent with observations of reduced noise (presumably due to fluctuations of charged impurities) in BLG compared to SLG[28]. Further work is needed to clarify that source of scattering in undoped BLG on $SiO_2$. Measurement of the variation of the mobility with dielectric constant[16] could potentially discriminate between charged-impurity and short-range disorder.

Acknowledgements

This work has been supported by the NSF-UMD-MRSEC grant DMR 05-20471 (E.D.W.) and the US ONR and ONR-MURI. The MRSEC shared equipment facilities were used in this work. Infrastructure support has also been provided by the UMD NanoCenter and CNAM. We thank S. Das Sarma for useful discussions regarding this manuscript.

Figure Captions

Figure 1.

FIG. 1 (a) Optical image of the bilayer graphene (BLG) device. Dark blue area is thick graphite ; red is the $SiO_2$/Si substrate, yellow areas are Cr/Au electrodes. The light purple rectangle is the bilayer graphene. (b) G' peak in micro-Raman spectrum acquired from the device area at 633 nm. Red line is fit to four Lorentzian components ; blue lines show the four individual components, numbers are relative offests of each Lorentzian in $cm^{-1}$.

Figure 2.

FIG. 2 The conductivity ($\sigma$) versus gate voltage ($V_g$) curves for different potassium concentrations for BLG (a) and SLG (b). For BLG, $\sigma(V_g)$ is measured at a temperature of 10K in UHV. Data in (b) are from Ref. [14].

Figure 3.

FIG. 3 (a) Inverse of electron mobility $1/\mu$ versus potassium concentration $n_K$. Line are linear fits to all data points used to extract the slope $1/C$. μ is the maximum field-effect mobility for SLG (data from Ref. [14]) and is shown at two different carrier densities for BLG. (b) The inverse slope $C$ from (a) versus effective gate voltage (solid black squares). Also shown is a second set of data from a different sample measured in a two-probe configuration (solid black circles). Solid lines show the theoretical predictions for $C$ within the Thomas-Fermi approximation for a parabolic dispersion relation assuming complete screening (black line) and finite TF screening wavevector with impurity graphene distance $d = 0$ (red) and $d = 0.43$ nm (purple). The green line shows the theoretical results for a hyberbolic dispersion relation with finite TF screening wavevector and $d = 0.43$ nm. The SLG value is also shown (blue dashed line) for comparison[14].

Figure 4.

FIG. 4 Minimum conductivity $\sigma_{min}$ and residual carrier density $n^*$ of bilayer graphene as a function of potassium concentration $n_K$. The blue (dashed) and black (solid) lines show fits to $n^* = [(n_{\mathrm{imp},0} + n_{\mathrm{K}})/\xi^2]^{1/2}$ and $\sigma_{min} = Ce[(n_{\mathrm{imp},0} + n_{\mathrm{K}})\xi^2]^{-1/2}$, with $C$ and $n_{\mathrm{imp},0}$determined from the fit to $1/\mu$ vs $n_{\mathrm{K}}$ at $V_{\mathrm{g}} = 30$ V in Figure 3a, and $\xi = 32$ nm.

Figure 1.

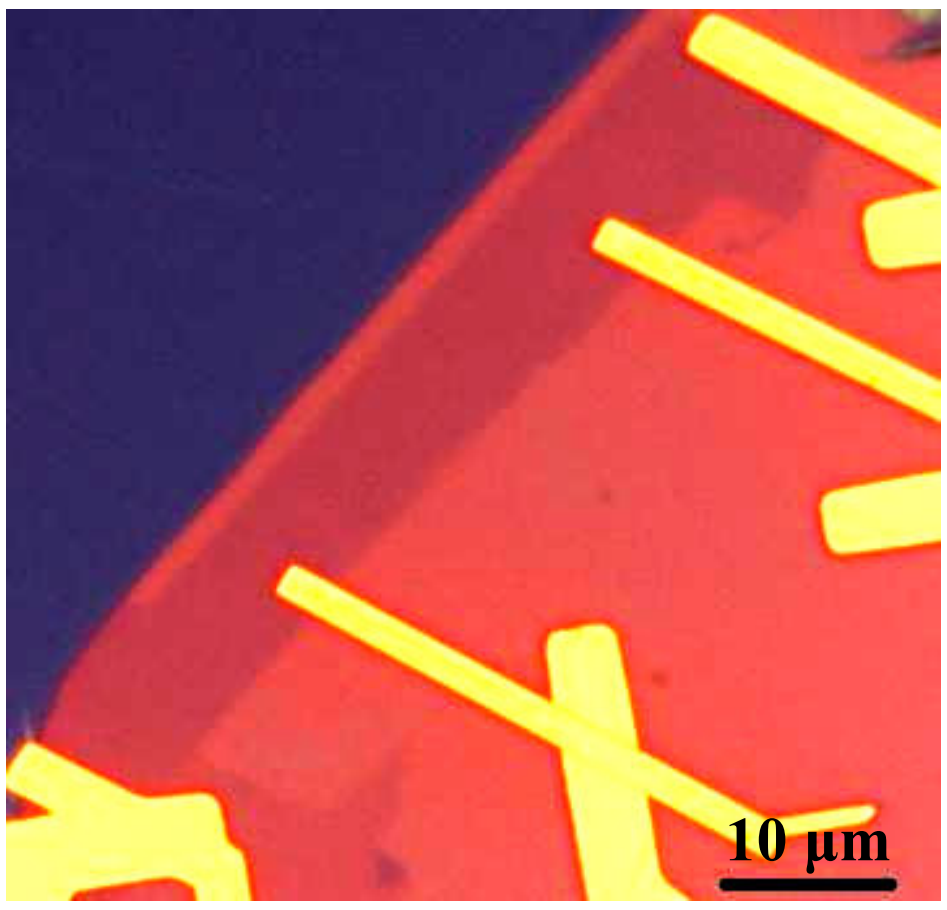
10 μm

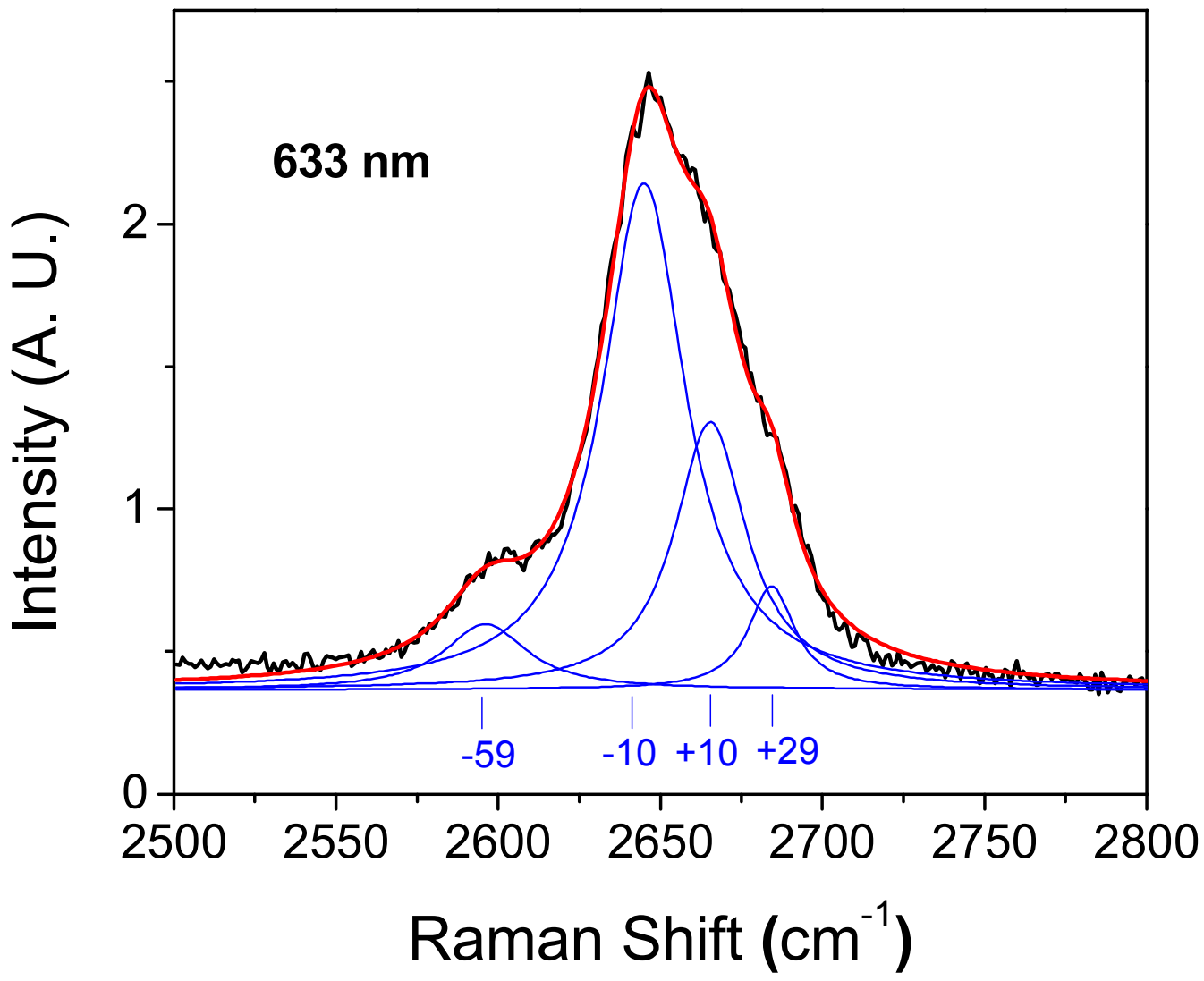

Figure 2.

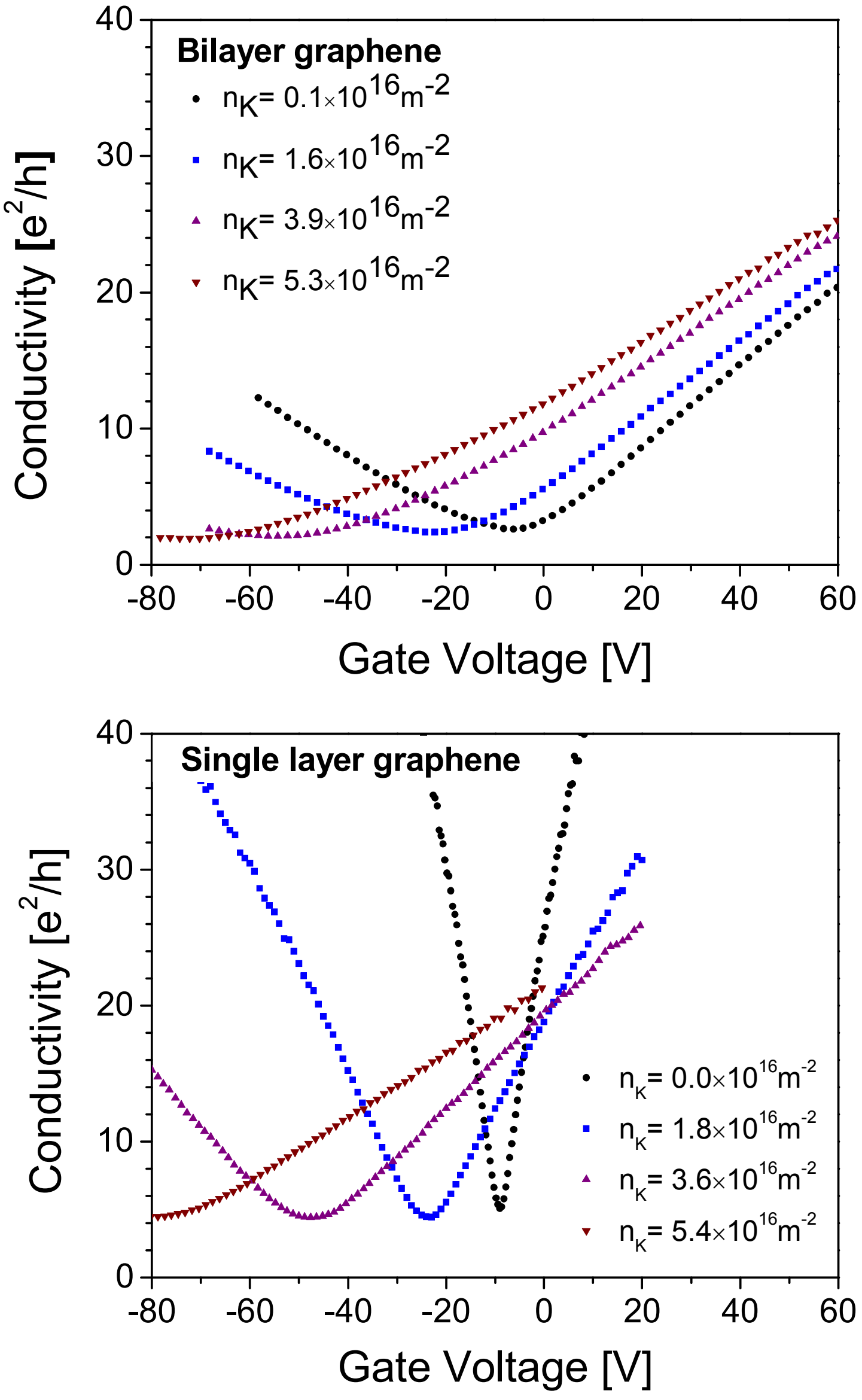

Figure 3.

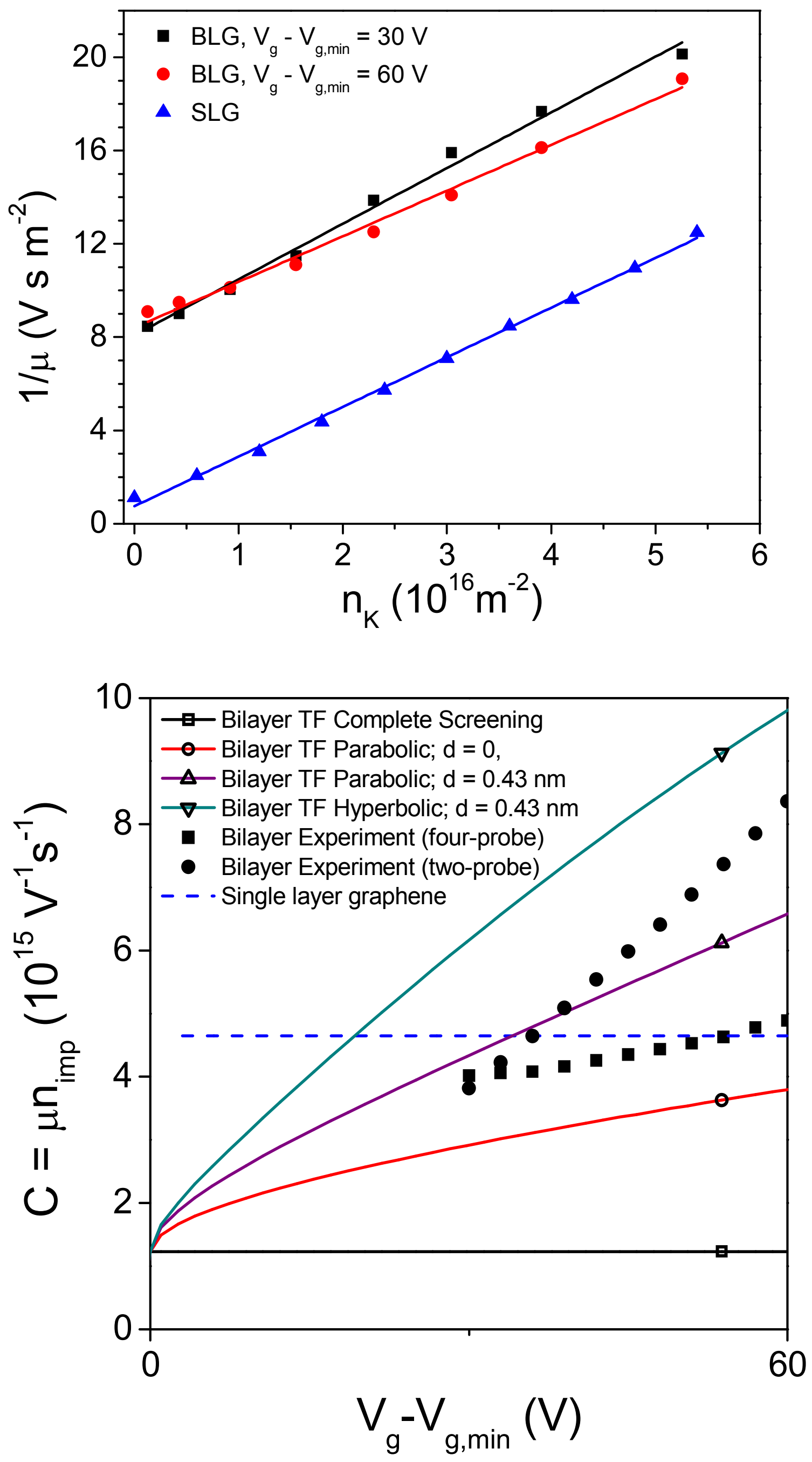
BLG, $V_g$ - $V_{g,min}$ = 30 V
BLG, $V_g$ - $V_{g,min}$ = 60 V
SLG
$1/\mu$ (V s $m^{-2}$)
$n_K$ ($10^{16} m^{-2}$)
Bilayer TF Complete Screening
Bilayer TF Parabolic; d = 0,
Bilayer TF Parabolic; d = 0.43 nm
Bilayer TF Hyperbolic; d = 0.43 nm
Bilayer Experiment (four-probe)
Bilayer Experiment (two-probe)
Single layer graphene
$C = \mu n_{imp}$ ($10^{15}$ $V^{-1}s^{-1}$)
$V_g$-$V_{g,min}$ (V)

Figure 4.

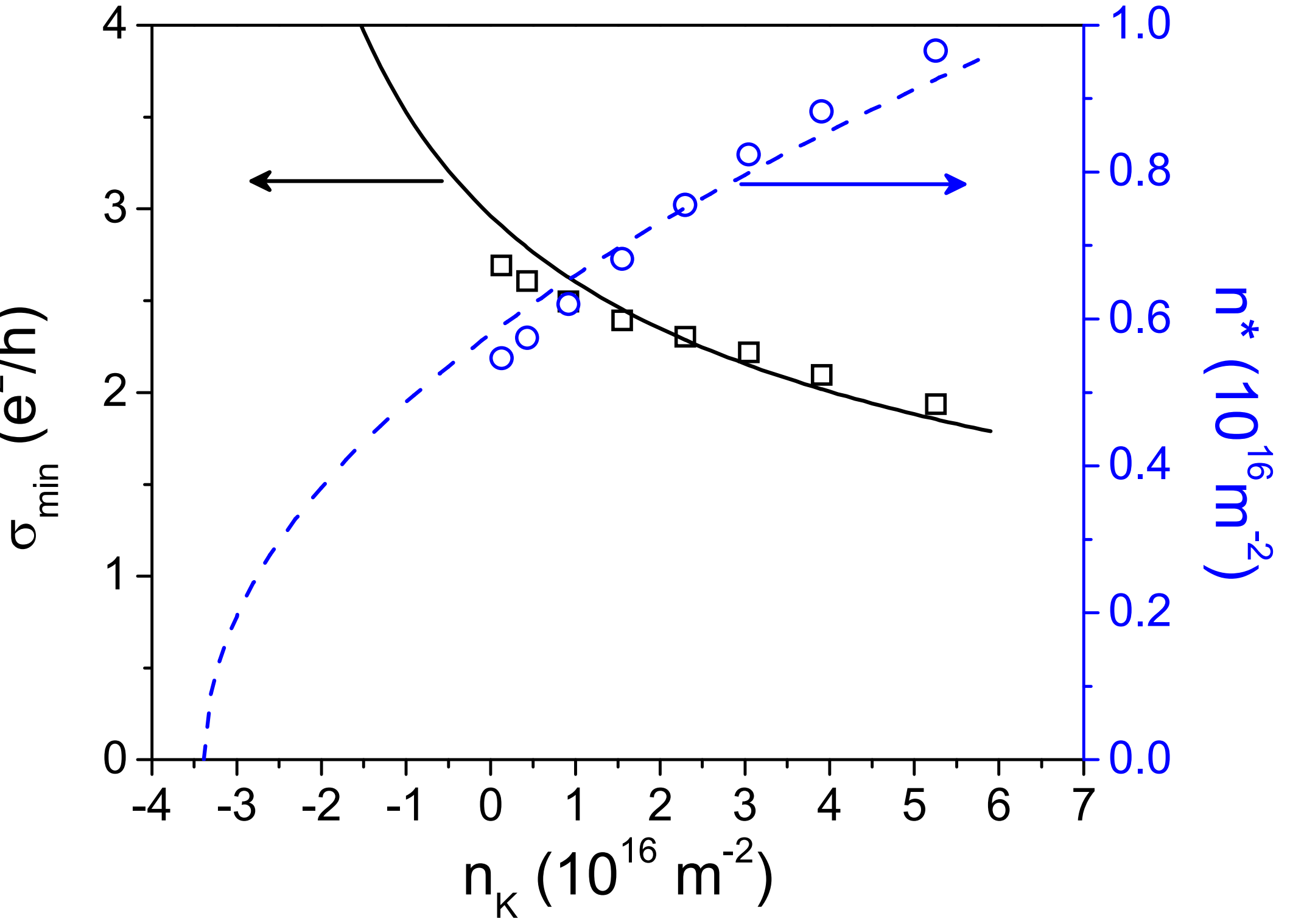